\documentclass[aps,pra,twocolumn,groupedaddress,amsmath,amssymb]{revtex4-1}
\usepackage{graphicx}
\usepackage{color}  % needed for figures
\usepackage{dcolumn}   % needed for some tables
\usepackage{bm}        % for math
\usepackage{verbatim}   % for math
\usepackage{hyperref}                   % for hyper refrences
\usepackage[autostyle]{csquotes}

\begin{document}

%\preprint{APS/123-QED}

\title{Relativistic Quantum Information of Anyons}% Force line breaks with \\
%\thanks{A footnote to the article title}%

\author{Leili Esmaeilifar}
   \email{l.esmaeilifar@iasbs.ac.ir}
   \author{Behrouz Mirza  }
  \email{b.mirza@cc.iut.ac.ir}
  \author{Hosein Mohammadzadeh}
  \email{mohammadzadeh@uma.ac.ir}

   \affiliation{*Department of Physics, Institute for Advanced Studies in Basic Sciences (IASBS), Zanjan 45137-66731, Iran }
   \affiliation{ $\dagger$ Department of Physics, Isfahan University of Technology, Isfahan 84156-83111, Iran}
 \affiliation{ $\dagger$ Department of Physics, University of Mohaghegh Ardabili, P.O. Box 179, Ardabil, Iran}

%\collaboration{CLEO Collaboration}%\noaffiliation

%%%\date{\today}% It is always \today, today,
             %  but any date may be explicitly specified

\begin{abstract}
\begin{center}Abstract:\\
\end{center}
 In this paper, a method is developed to investigate the relativistic quantum information of anyons.
Anyons are particles with intermediate statistics ranging between Bose-Einstein and Fermi-Dirac
statistics, with a parameter $\alpha$ ($0<\alpha<1$)
characteristic of this intermediate statistics. A density matrix is also introduced as a combination of the density matrices of bosons and fermions with a continuous parameter,
$\alpha$, that represents the behavior of anyons. This density matrix reduces to bosonic and fermionic density matrices in the limits $\alpha\rightarrow 0$ and $\alpha\rightarrow 1$,
 respectively. We compute entanglement entropy, negativity, and coherency for anyons in non-inertial frames as a function of $\alpha$. We also computed quantum fisher information for these particles. Semions, which are particles with $\alpha = 0.5$, were
found to have minimum quantum fisher information with respect to $\alpha$ than those with other values of fractional parameter.
\end{abstract}

%%%\pacs{Valid PACS appear here}% PACS, the Physics and Astronomy
                             % Classification Scheme.
%\keywords{Suggested keywords}%Use showkeys class option if keyword
                              %display desired
\maketitle

%\tableofcontents

\section{\label{sec:level1}Introduction}
Particles in the three-dimensional or higher space are classified, based on their statistical behavior, as bosons and fermions.
The multi-particle wave function of identical bosons (fermions) is symmetric (antisymmetric) under interchange of any pair of particles.
It has been shown that quasiparticles in the two-dimensional space may have intermediate statistics between bosons and fermions with a continuous parameter.
 This can be written in the  two particles' case as follows:
\begin{equation}
\mid \psi_{1}\psi_{2}\rangle = e^{i \pi \alpha}\mid \psi_{2}\psi_{1}\rangle,
\end{equation}
where, $\alpha$ is the fractional statistical exchange quantum number, also called the statistical parameter, ranging over $0\leq \alpha\leq1$,
with $\alpha = 0$ standing for bosons and $\alpha = 1$ for fermions. The theoretical possibility of these particles was first propounded by J. M. Leinaas and J. Myrheim \cite{leinaas1977theory}.
They later came to be called anyon by F. Wilczek \cite{wilczek1982quantum}.

The Pauli exclusion principle was generalized to yield another sort of generalized statistics introduced by F. D. M. Haldane \cite{haldane1991fractional}.
This generalization is independent of the dimension of the system. A fractional parameter, $g$, was defined for the fractional exclusion statistics. In this exclusion statistics two limits are defined  for  g, where  g = 0 (g=1)  corresponds to bosons (fermions). Despite the radical differences in the basic definitions of the fractional exchange and fractional exclusion statistics, the relationship between these fractional statistics has been investigated in the two-dimensional space. Particles with a fractional statistics in the two-dimensional space are sometimes called anyons. For our purposes in this study, we will consider the Haldane fractional exclusion statistics in two dimensions and use anyon to call a particle.
\\ Using Haldane’s fraction exclusion statistics, Wu derived the statistical distribution function of anyons as follows \cite{wu1994statistical}:
\begin{equation}
n_{i}=\frac{1}{\omega(\textmd{e}^{(\epsilon_{i}-\mu)/K T})+\alpha}.
\end{equation}
The functional equation of $\omega$ is $\omega(x)^{\alpha}[1+\omega(x)]^{(1-\alpha)}=x$.
\\ This is an active area of research for its important role in such different fields as quantum computations \cite{nayak2008non} and fractional quantum Hall effect \cite{stern2008anyons}.

There exist a factorizable property for the thermodynamic quantities of a two dimensional gas with particles which obeying the fractional exclusion statistics. In fact, it has been shown that the system with bosons and fermions, by allowing a transmutation between them will have the statistical distribution function of fractional exclusion statistics which called anyon in two dimension \cite{huang1995,huang1996statistics}. This property motivate us to construct an appropriate density matrix for anyons.

 Recently, quantum information in a relativistic limit and in non-inertial frames has attracted the attention of many researchers to the new field of relativistic quantum information \cite{fuentes2005alice,alsing2006entanglement,mehri2011pseudo,mehri2015quantum,fuentes2010entanglement,mohammadzadeh2015entanglement,
mohammadzadeh2017entropy,friis2013entanglement}.
Relativistic quantum information (RQI) studies the relationship between quantum information theory and special and general relativity.  This field has some applications in fundamental physics such as cosmology, black hole physics, and some approaches to quantum gravity\cite{cosmo1,cosmo2,bh1,bh2}. Besides, there are also lots of applications of RQI in the newfound techniques, in different areas of quantum information theory like quantum communication, quantum computing, and quantum metrology \cite{qcomu1,qcomu2,qcomu3}.  In all of these areas, entanglement and quantum correlations play an important role\cite{Nielsen,wilde2013}.  specifically, in the case of satellite-based quantum communication,  general and special relativistic effects could change the efficiency of communication \cite{satellite-basedqi}. Therefore, it is important to investigate the effects of relativity on entanglement and other quantum correlations.
\\ The entanglement of bosons and fermions have been studied with interesting results \cite{fuentes2005alice,alsing2006entanglement}.
 It has been found that entanglement degrades at high limit accelerations in both bosonic and fermionic cases as a result of appearance of a horizon in accelerated frames.
 Entanglement in bosonic modes has been found to vanish but that of fermionic modes to survive at the infinite limit of acceleration.
Researchers have also investigated entanglement generation for boson and fermion modes in an expanding universe \cite{fuentes2010entanglement,mohammadzadeh2015entanglement,liu2016quantum}.

Quantum teleportation in accelerated frames and in the background of Schwarzschild spacetime \cite{mehri2015quantum,xiao2020enhanced}, densecoding in non-inertial frames \cite{farahmand2017superdense,grochowski2017effect} and transmission of quantum information through quantum fields \cite{simidzija2020transmission} are some examples of quantum information processes in relativistic domain.

 Moreover, investigations have shown behavioral differences between boson and fermion modes in relativistic frames.
 The present study considers anyon modes in a non-inertial frame and investigates variations in entanglement with respect to the acceleration as an attempt to shed more light on the differences between boson and fermion modes. To achieve this goal, a model is introduced for the study of entanglement of anyons in non-inertial frames. The results thus obtained will be compared with those obtained for bosons and fermions in non-inertial frames.

The paper comprises the following four sections. In Section \ref{sec:level2}, a brief review is presented of previous studies of entanglement of fermion and boson modes in non-inertial frames.
In Section \ref{sec:level3}, a density matrix is proposed for anyon modes and entanglement variation in response to varying accelerations is studied for different values of the fractional parameter. In \ref{sec:level4} relative entropy of coherence is computed. Quantum fisher information is studied in sec \ref{sec:level5}.
Finally, a summary of the results is presented in Section \ref{sec:level6}.

\section{\label{sec:level2}Entanglement Entropy and Negativity of bosons and fermions}

\begin{figure}[h]
\centering
\includegraphics[width=0.4\textwidth]{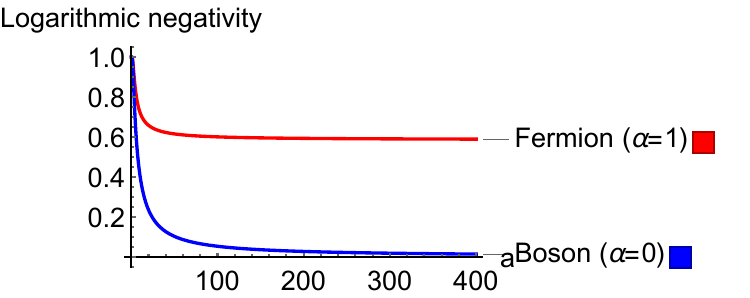}
\caption[]{\it Negativity as a function of acceleration for bosons and fermions when $\theta=\frac{\pi}{4}$  } \label{fig:1}
\end{figure}

We consider an inertial observer, named Alice, who has a detector sensitive to modes $k_{A}$.
Another observer, named Rob who moves with a uniform acceleration $(a)$, has a detector sensitive to modes $k_{R}$.
Then, we consider an entangled Bell state for the two maximally fermionic modes, $ k_{\textmd{A}}$ and $k_{\textmd{R}}$ as follows:
\begin{equation}
 \label {eq3}
\mid \psi _{k_{\textmd{A}},k_{\textmd{R}}}\rangle =\textmd{cos}(\theta)\mid 0_{k_{\textmd{A}}}\rangle^{+}\mid 0_{k_{\textmd{R}}}\rangle^{+}+\textmd{e}^{i\phi} \textmd{sin}(\theta)\mid 1_{k_{\textmd{A}}}\rangle^{+}\mid 1_{k_{\textmd{R}}}\rangle^{+},
\end{equation}
where, $+$ is used to show the positive answers of Dirac fields (particles) and $\theta$ and $\phi$ are called weight and phase parameters respectively.
We write the expansion of $\mid 0_{k_{R}}\rangle$ and $\mid 1_{k_{R}}\rangle$ in Rob's and antiRob's states in Rindler coordinates, which are in
two different causally disconnected regions, $I$ and $II$, respectively.
Finally, the reduced density matrix for the entangled state observed by Alice and Rob $(\rho\textmd{(A,I)=Tr}_{\textmd{II}}(\mid\psi\rangle\langle\psi\mid))$ for fermions is found as follows \cite{alsing2006entanglement}:
\begin{eqnarray}\label {eq:4}
\nonumber\rho_{\textmd{f}}\textmd{\textmd{(A,I)}}&=& \textmd{cos}(\theta) \textmd{cos}(\gamma)\mid 0,0\rangle \langle 0,0\mid
 \\\nonumber &&+\textmd{sin}^{2}(\gamma)\textmd{cos}^{2}(\theta)\mid 0,1\rangle \langle 0,1\mid  +\textmd{sin}^{2}(\theta)\mid 1,1\rangle \langle 1,1\mid \\ &&+\frac{\textmd{sin}(2\theta)}{2} \textmd{cos}(\gamma)(\textmd{e}^{-i\phi}\mid 0,0\rangle \langle 1,1\mid +H.c.) ,
\end{eqnarray}
where, $\textmd{tan}(\gamma) =\textmd{exp}(\frac{\pi \omega_{\textmd{f}}}{a})$, $\omega_{\textmd{f}}$ is the frequency of the fermionic modes detected by Alice and Rob, $a$ is the relative acceleration, and $\mid a,b\rangle=\mid a\rangle_{\textmd{A}}\mid b\rangle_{\textmd{I}}$.
One can construct the matrix form on the basis of $\mid0,0\rangle ,\mid0,1\rangle,\mid 1,0\rangle,\mid 1,1\rangle$, which has the eigenvalues $\{0,0,1-\textmd{sin}^{2}(\gamma) \textmd{cos}^{2}(\theta) ,\textmd{sin}^{2}(\gamma) \textmd{cos}^{2}(\theta)\}$.
The density matrix for the bosonic case ($\rho_{\textmd{b}}$) is obtained as follows \cite{fuentes2005alice}:
\begin{equation}\label {eq:5}
\rho_{\textmd{b}}\textmd{(A,I)}=\frac{1}{\textmd{cosh}^{2} (r)}\sum_{\textmd{n}=0}^{ \infty}\textmd{tanh}^{2\textmd{n}}(r) \rho_{\textmd{b}}^{(\textmd{n})},
\end{equation}
where, $\textmd{tanh}(r)=\textmd{exp}(\frac{\pi \omega_{b}}{a})$ and again $a$ is the relative acceleration , $\omega_{b}$ is the frequency of the bosonic modes and $\rho_{b}^{(\textmd{n})}$ is defined as follows:
\begin{eqnarray}
\rho_{b}^{(\textmd{n})}&& =\textmd{cos}^{2}(\theta)\mid 0_{k},\textmd{n}_{k'}\rangle \langle 0_{k},\textmd{n}_{k'}\mid +\\\nonumber && (\textmd{n}+1)\frac{\textmd{sin}^{2}(\theta)}{\textmd{cosh}^{2}(r)}\mid 1_{k},(\textmd{n}+1)_{k'}\rangle\langle 1_{k},(\textmd{n}+1)_{k'}\mid+\\\nonumber &&\sqrt{\textmd{n}+1}\frac{\textmd{cos}(\theta)\textmd{sin}(\theta)}{\textmd{cosh}(r)}\Big(\textmd{e}^{i\phi} \mid 0_{k},\textmd{n}_{k'}\rangle \langle 1_{k},(\textmd{n}+1)_{k'}\mid +H.c.\Big).
\end{eqnarray}
Although $\rho_{\textmd{b}}$ is infinite-dimensional, it is block diagonal ($\rho^{(\textmd{n})}_{b}$), allowing it to be diagonalized block by block, and its eigenvalues to be found.
To quantify the entanglement of this system, we compute the logarithmic negativity \cite{vidal2002computable} defined as follows:
\begin{eqnarray}
E_{N}(\rho)& =&\log_{2}(1+2\sum_{i}\frac{\mid \lambda_{i}(\rho^{PT_{\textmd{A}}})\mid -\lambda_{i}}{2}),
 \end{eqnarray}
 where, $ \lambda_{i}(\rho^{PT_{\textmd{A}}})$ is the eigenvalues of partial transpose of $\rho\textmd{(A,I)}$ obtained by exchanging Alice's qubits.
Negativity (or logarithmic negativity) measures the entanglement of a system in a way that the system is not entangled for a negativity equal to zero.
The logarithmic negativities for bosons and fermions for the maximally entangled case ($\theta=\frac{1}{\sqrt{2}}, \phi=0$) are plotted in Fig.(\ref{fig:1}) (In this paper we assume that $\omega_{\textmd{f}}=\omega_{\textmd{b}}=1$).

%%%%%%%%%%%%%%%%%%%%%%%%%%%%%%%%%%%%%%%%%%%%%%%%%%%%%
\section{\label{sec:level3}Entanglements of anyon modes}
%%%%%%%%%%%%%%%%%%%%%%%%%%%%%%%%%%%%%%%%%%%%%%%%%%%%%
As mentioned in the introduction, anyons are particles with intermediate statistics. It has been shown that a system of anyons in the two-dimensional space can be considered as
the ensemble average of bosons and fermions with the fractions of $\alpha$ and $(1-\alpha)$, respectively. This property stems from
the fact that the density of states is constant in a two-dimensional space.
Thus, the ensemble average of any thermodynamic quantity, like the internal energy or particle number, can be factorized as follows:
\begin{equation}\label{e8}
Q(\alpha)=\alpha Q_{\textmd{f}}+(1-\alpha)Q_{\textmd{b}},
\end{equation}
where, $Q(\alpha)$ denotes the thermodynamic quantity of anyons and $Q_{f}$ and $Q_{b}$ refer to those of fermions and bosons, respectively \cite{huang1996statistics,mirza2009nonperturbative,PhysRevE.82.031137,b}. We may note that the thermodynamics of a system can be obtained from the partition
function or, equivalently, from the density matrix.
Exploiting this idea, we introduce a new density matrix as a direct sum of boson's
and fermion's density matrices with a continuous parameter $\alpha$, called the statistical parameter as follows:
 \begin{equation}\label{roanyon}
 \rho_{\textmd{a}}\textmd{(A,I)} =(1-\alpha) \rho_{\textmd{b}}\textmd{(A,I)} \oplus  \alpha \rho_{\textmd{f}}\textmd{(A,I)}.
 \end{equation}
This density matrix mimics the behavior of anyons in non-inertial frames.
 $\rho_{\textmd{a}}$ is a block diagonal matrix, and its eigenvalues $\lambda_{i*j}^{(\textmd{n})}$ are those of $\rho_{\textmd{f}}$ and $\rho_{\textmd{b}}$
\begin{equation}
 \lambda_{i*j}(\rho_{a})=\lbrace (1-\alpha) \lambda _{i}(\rho_{\textmd{b}}), \alpha\lambda _{j}(\rho_{\textmd{f}}) \rbrace .
\end{equation}
The entanglement entropy of $\rho_{\textmd{a}}$ is as follows:
\begin{eqnarray}\label{sa}
S(\rho_{\textmd{a}})&=&\sum_{i}\lambda_{i}(\rho_{\textmd{a}}) Log_{2}(\lambda_{i}(\rho_{\textmd{a}}))\\\nonumber &&= S((1-\alpha)\rho_{\textmd{b}})+ S(\alpha \rho_{\textmd{f}})) ,
\end{eqnarray}

\begin{figure}[h]
\centering
\includegraphics[width=0.4\textwidth]{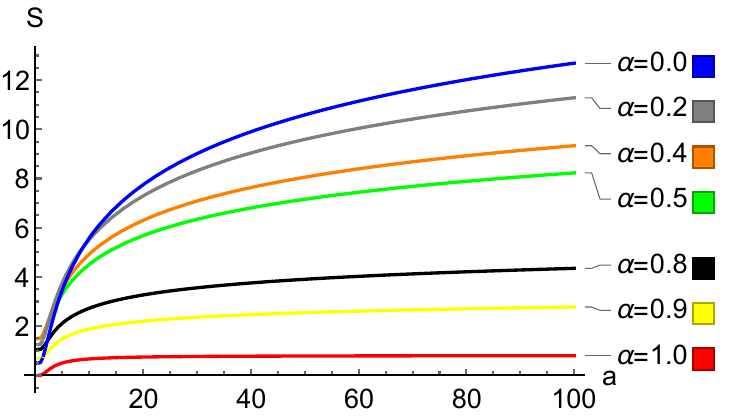}
\caption[]{\it Entanglement entropy of anyons as a function of acceleration for different values of  $\alpha$ when $\theta=\frac{\pi}{4}$.} \label{fig:2}
\end{figure}

Entanglement entropy of anyons is plotted in Fig.(\ref{fig:2}).
It is easy to check that $\alpha=0$ and $\alpha=1$ in Fig.(\ref{fig:2}) are the bosonic and fermionic cases.

Since the entanglement entropy is not a suitable measure for
mixed states \cite{27}, other measures, like negativity, may be employed to
explore the entanglement of the system.
%%%%%%%%%%%%%%%%%%%%%%%%%%%%%%%%%%%%%%%%%%%%
\\To compute the logarithmic negativity, we need to find the partial transpose of  $\rho_{a}$ which is defined as below:
\begin{equation}\label{eq17}
 \rho_{a}^{\textmd{PT}}\textmd{(A,I)} =(1-\alpha) \rho_{\textmd{b}}^{\textmd{PT}} \oplus \alpha \rho^{\textmd{PT}}_{\textmd{f}}.
\end{equation}
Therefore, the negativity of anyons is :
\begin{eqnarray}
N(\rho)&& = \sum_{k}\frac{\mid \lambda_{k}(\rho^{\textmd{PT}}_{a})\mid -\lambda_{k}(\rho^{\textmd{PT}}_{a})}{2} \\\nonumber &&=(1-\alpha) N(\rho_{b})+ \alpha N(\rho_{f}).
\end{eqnarray}

\begin{figure}[h]
\centering
\includegraphics[width=0.4\textwidth]{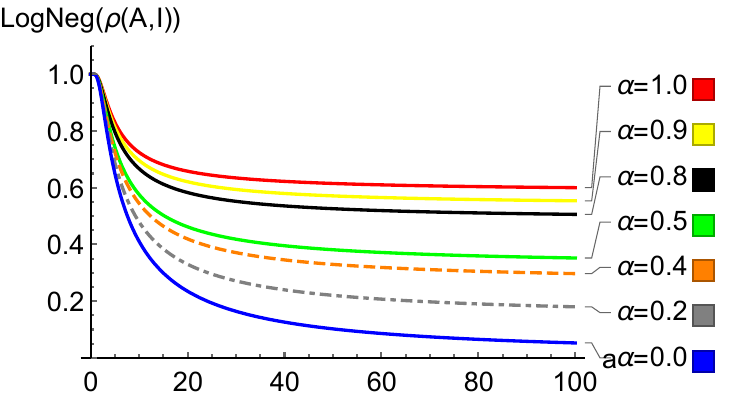}
\caption[]{\it Logarithmic negativity of anyons as a function of acceleration for different values of statistic parameter when $\theta=\frac{\pi}{4}, \phi=0$. Solid red curve: $\alpha =1$, Solid yellow curve: $\alpha =0.9$. Solid black curve: $ \alpha =0.8$. Solid green curve: $\alpha =0.5$. Dashed orange curve: $\alpha =0.4$. DotDashed gray curve: $\alpha =0.2$. Solid blue curve: $\alpha =0$.} \label{fig:3}
\end{figure}

This is plotted for different values of $\alpha$ in Fig.(\ref{fig:3}).

As $\alpha$ tends to $0$, negativity declines with increasing acceleration. Ultimately, in the bosonic case of $\alpha=0$, negativity becomes $0$ in the limit $a\rightarrow \infty$.

\section{\label{sec:level4}Coherency}
\begin{figure}[h]
\centering
\includegraphics[width=0.4\textwidth]{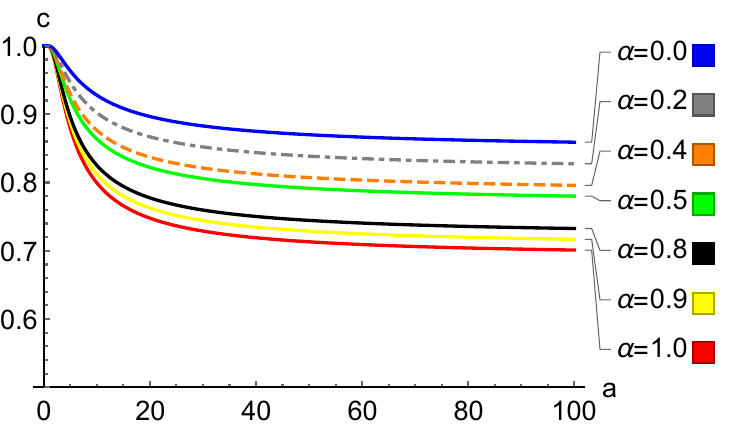}
\caption[]{\it Relative entropy of coherence of anyons as a function of acceleration for different values $\alpha$ when $\theta=\frac{\pi}{4}, \phi=0$.} \label{fig:4}
\end{figure}

Quantum coherence forms another useful subject in quantifying quantum correlations.
Unlike entanglement which is used for an interacting system, quantum coherence is of interest in systems with no interaction.
Coherence measures more quantum correlations than entanglement does. From among the coherence quantifiers available, we use the relative entropy of coherence defined as follows: \cite{baumgratz2014quantifying}
\begin{equation}\label{c}
C_{r}=S(\rho_{\textmd{diagonal}})-S(\rho),
\end{equation}
where, $\rho_{diagonal}$ is obtained by deleting the off-diagonal elements of $\rho.$ Using Eqs.(\ref{roanyon}) and (\ref{sa}), we can compute the relative entropy of coherence (Eq.(\ref{c})). Like the other measures, the relative entropy of coherence is also the summation of boson and fermion relative entropy of coherence with the coefficients $(1-\alpha)$ and $\alpha$, respectively.
\\In Fig.(\ref{fig:4}), the relative entropy of coherence of this system is plotted for different values of the statistical parameter.\\
As $\alpha$ distances more away from $1$, $C_{r}$ rises to reach its maximum at $\alpha=0$ (bosons). Correlations for all values of statistical parameters never vanish.

\section{\label{sec:level5} quantum fisher information (QFI)}
\begin{figure}[h]
\centering
\includegraphics[width=0.4\textwidth]{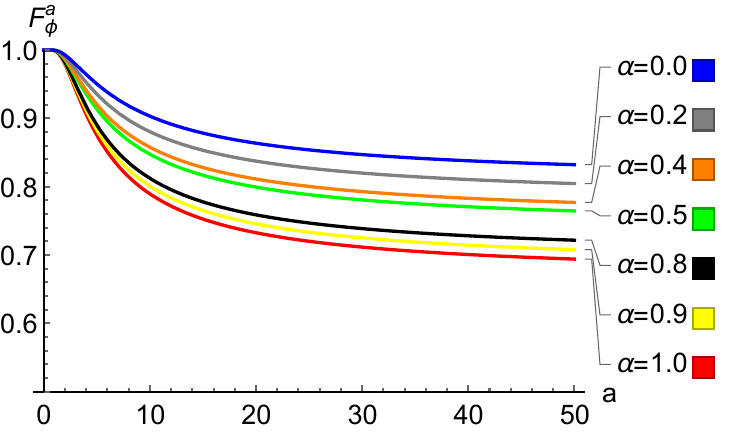}
\caption[]{\it  Quantyum fisher information of anyons with respect to $\phi$ when $\theta=\frac{\pi}{4}$ as a function of acceleration for different values $\alpha$.} \label{fig:5}
\end{figure}

Parameter estimation of a probability distribution has a central role in the field of quantum information. Assuming a distribution, $X$, that models an observable $\hat{O}$,  we may define fisher information as a measure to identify the information that $\hat{O}$  has about a parameter of $X$ \cite{fisher1,fisher2}.  Since the quantum measurement is essentially probabilistic, quantum fisher information has been introduced. The quantum fisher information can be used to extract
information about an unknown parameter $q$ form the density matrix, $\rho(q)=\sum_{i}\lambda_{i}(q)\mid \Phi_{i}(q)\rangle\langle\Phi_{i}(q)\mid$. QFI is defined as follows \cite{fisher}:
 \begin{equation}
 F_{q}=\sum _{i} \frac{(\lambda _{i}^{'})^{2}}{\lambda _{i}}+ \sum _{i} \lambda_{i} F_{q,i} -\sum_{i\neq j} \frac{8 \lambda_{i}\lambda_{j}}{\lambda_{i}+\lambda_{j}}\mid \langle\Phi_{i}\mid\Phi_{j}\rangle|^{2},
 \end{equation}
where, $\lambda _{i}^{'}=\partial_{q}\lambda _{i}$ and $F_{q,i}$ is defined as follows:

\begin{equation}
F_{q,i}=4 (\langle\partial_{q} \Phi_{i}\mid \partial_{q} \Phi_{i}\rangle -\mid \langle \Phi_{i}\mid \partial_{q} \Phi_{i}\rangle \mid^{2}).
\end{equation}
The first term of $ F_{q}$ is the classical part and the second term is the quantum parts of QFI. The third term
arises from the mixture of pure states. Density matrix of fermion in Eq.(\ref{eq:4}) can be written as follows :
   \begin{equation}
 \rho_{\textmd{f}}\textmd{(A,I)} = \lambda_{1}^{\textmd{f}} \mid \Phi_{1}^{\textmd{f}}\rangle\langle\Phi_{1}^{\textmd{f}}\mid+ \lambda_{2}^{\textmd{f}}  \mid \Phi_{2}^{\textmd{f}}\rangle\langle\Phi_{2}^{\textmd{f}}\mid,
  \end{equation}
where, $\lambda_{i}$ are the nonzero eigenvalues of $\rho\textmd{(A,I)}^{\textmd{f}}$
\begin{eqnarray}
\lambda_{1}^{\textmd{f}}&=&1-\textmd{sin}^{2}(\gamma)\textmd{cos}^{2}(\theta),\\\nonumber \lambda_{2}^{f}&=&\textmd{sin}^{2}(\gamma)\textmd{cos}^{2}(\theta),
  \end{eqnarray}
  and the  $\mid \Phi_{i}^{f}\rangle $ are the normalized eigenvectors of $\rho_{\textmd{f}}\textmd{(A,I)}$
   \begin{eqnarray}
 &&\mid \Phi_{1}^{\textmd{f}}\rangle =\frac{1}{\sqrt{1+\textmd{cos}^{2}(\gamma)\textmd{cot}^{2}(\theta)}}\lbrace e^{-i\phi}\textmd{cos}(\gamma)\textmd{cot}(\theta),0,0,1\rbrace ,
  \nonumber\\&&\mid \Phi_{2}\rangle =\lbrace 0,1,0,0\rbrace.
  \end{eqnarray}
 For the bosonic case we have
 \begin{equation}
\rho_{\textmd{b}}\textmd{(A,I)}= \sum_{\textmd{n}=0}^{\infty}\lambda_{\textmd{n}}^{\textmd{b}}\mid\Phi^{\textmd{b}}_{\textmd{n}}\rangle\langle\Phi^{\textmd{b}}_{\textmd{n}}\mid,
\end{equation}
where,
\begin{eqnarray}
\mid \Phi_{\textmd{n}}^{\textmd{b}}\rangle&=&\{\frac{\textmd{cos}(\theta)}{\sqrt{\vartheta_{\textmd{n}}}},\frac{e^{i\phi}\sqrt{\textmd{n}+1}\textmd{sin}(\theta)}{\textmd{cosh}(r)\sqrt{\vartheta_{\textmd{n}}}}\rbrace^{T},\\\nonumber &
\lambda_{\textmd{n}}^{\textmd{b}}&=\frac{\textmd{tanh}^{2\textmd{n}}(r)}{\textmd{cosh}^{2}(r)}\vartheta_{\textmd{n}},\\\nonumber &
 \vartheta_{\textmd{n}}&=\textmd{cos}(\theta)+\frac{\sqrt{\textmd{n}+1}\textmd{sin}(\theta)}{\textmd{cosh}(r)}\nonumber.
\end{eqnarray}
 It can be seen that the QFI with respect to $\theta$ for both fermions and bosons is equal to 4. However, $F_{\phi}$ for fermions and bosons has functions as below \cite{fisherNI}:
 \begin{eqnarray}
 F_{\phi}^{\textmd{b}}&=&\frac{\textmd{sin}^{2}(2\theta)}{\textmd{cosh}^{2}(r)}\sum_{\textmd{n}=0}^{\infty}\frac{(\textmd{n}+1)\textmd{tanh}^{2\textmd{n}}(r)}{\vartheta_{\textmd{n}}},\nonumber\\ &&
 F_{\phi}^{\textmd{f}}=\frac{\textmd{sin}^{2}(2\theta)\textmd{cos}^{2}(\gamma)}{1-\textmd{sin}^{2}(\gamma)\textmd{cos}^{2}(\theta)}.
 \end{eqnarray}
%According to Eq.(\ref{roanyon}), the density matrix of anyon can be written as
 %\begin{eqnarray}
 %\rho_{\textmd{a}}\textmd{(A,I)}&=&\sum_{i}\lambda^{\textmd{a}}_{i}\mid\Phi^{\textmd{a}}_{i}\rangle\langle\Phi^{\textmd{a}}_{i}\mid \\\nonumber&
 %=&(1-\alpha)\sum_{\textmd{n}=0}^{\infty}\lambda_{\textmd{n}}^{\textmd{b}}\mid\Phi^{\textmd{b}}_{\textmd{n}}\rangle\langle\Phi^{\textmd{b}}_{\textmd{n}}\mid +\\\nonumber &&
 %\alpha\Big(\lambda_{1}^{\textmd{f}} \mid \Phi_{1}^{\textmd{f}}\rangle\langle\Phi_{1}^{\textmd{f}}\mid+ \lambda_{2}^{\textmd{f}} \mid \Phi_{2}^{\textmd{f}}\rangle\langle\Phi_{2}^{\textmd{f}}\mid\Big).
 %\end{eqnarray}
 Therefore, it is obvious that $F_{\theta}^{\textmd{a}}=4$ and $F_{\phi}^{\textmd{a}}= \alpha \textmd{F}_{\phi}^{\textmd{f}}+(1-\alpha)\textmd{F}_{\phi}^{\textmd{b}}$ (Fig.(\ref{fig:5})).
For the computation of $F_{\alpha}^{\textmd{a}}$ we only need to compute the classical part of QFI, since only $\lambda_{i}^{\textmd{a}}$ depends on $\alpha$. and the result is as follows
\begin{equation}
F_{\alpha}^{a}=\frac{1}{\alpha(1-\alpha)}.
\end{equation}
For semions, particles with $\alpha=0.5$, QFI is minimum, while for bosons and fermions it tends to infinity.
\section{\label{sec:level6}Summery and conclusion}
We proposed a density matrix for anyon modes as a linear combination of the density matrices of boson and fermion modes (Eq.(\ref{roanyon})).
%determine the entanglement between two free anyonic modes by each detecting one of the modes.
We found the entanglement entropy of the system and compared it with those of the bosonic and fermionic cases.
The entanglement between  anyon modes was investigated by computing the logarithmic negativity of the system.
 In the bosonic limit, $\alpha=0$, negativity was observed to vanish for $a\rightarrow \infty$, as expected. To consider all the possible correlations, we also computed the relative
entropy of coherence to find that anyons are always coherent for all values of both acceleration and statistical parameter. In addition we computed the quantum fisher information with respect to $\theta$, $\phi$ and statistical parameter $\alpha$. Also the QFI with respect to statistical parameter for semions is minimum.
It is interesting to further explore relativistic quantum information of anyons in other curved spacetimes \cite{anyonblackhole}.

\end{document}